# CHIRAL SYMMETRY AND LATTICE GAUGE THEORY[*]

MICHAEL CREUTZ

*Physics Department, Brookhaven National Laboratory*
*Upton, NY 11973, USA*

## ABSTRACT

I review the problem of formulating chiral symmetry in lattice gauge theory. I discuss recent approaches involving an infinite tower of additional heavy states to absorb Fermion doublers. For hadronic physics this provides a natural scheme for taking quark masses to zero without requiring a precise tuning of parameters. A mirror Fermion variation provides a possible way of extending the picture to chirally coupled light Fermions.

## 1. Introduction

Lattice gauge theory is now entering its third decade. Throughout this period Fermions have presented challenging obstacles. One of these issues lies in Monte Carlo algorithms, which remain awkward when quarks are included. Indeed, in the presence of a background Fermion density representing a chemical potential, no truly viable algorithms are known.

In this talk, however, I will concentrate on the other "Fermion problem," that of doubling and chiral symmetry. This discussion is essentially an abridged version of [1]. For a more general review of the subject see Ref. [3].

Why do we care about chiral symmetry when the lattice gives us a first principles scheme for calculating hadronic physics? The reason is partly historical and partly aesthetic. Indeed, chiral symmetry has long played an essential role in particle theory. The pion is made of the same quarks as the rho meson, yet its mass is considerably less. The canonical explanation says that if quarks were massless, then the pions would be Goldstone bosons arising from the spontaneous breaking of an underlying chiral symmetry. Naively, pure gauge interactions are helicity conserving, and thus both the number of left and right handed massless quarks are separately conserved. Through confinement into the physical states of baryons and mesons, this symmetry is spontaneously broken. A host of predictions from current algebra are based on this picture [2].

These issues are complicated by the presence of "anomalies." Ultraviolet divergences make it impossible, even in perturbation theory, to conserve simultaneously all axial vector currents associated with chiral symmetry, and the vector currents coupled to gluons. As current conservation is crucial to our understanding of gauge symmetries, we must conserve the vector currents, implying that the axial symmetry cannot be exact.

One consequence of the anomaly is that there is one less Goldstone boson than naive counting would suggest. For two flavors of light quarks, of the four ways to form pseudoscalar mesons there are three light pions while the eta remains heavier. With $SU(3)$ flavor symmetry, it is the $\eta'$ which is anomalously heavy compared to the other pseudoscalars.

Chiral issues arise in an even more fundamental way with the weak interactions. Here parity violation seems to maximally differentiate between left and right handed Fermions. While lattice methods have been dominantly applied to the strong interactions, there are reasons to desire a lattice formulation of the weak interactions as well. In particular, the lattice is the best founded non-perturbative regulator, and thus provides an elegant framework for the definition of a quantum field theory. Even though the smallness of the electromagnetic coupling makes non-perturbative effects quite small in the electroweak theory, at least in principle we would like a rigorous formulation. While exceptionally small, some interesting non-perturbative phenomena are directly related to the anomaly, such as the prediction that baryons can decay through the instanton mechanism of 't Hooft [4].

## 2. Massless Fermions and the anomaly

Chiral symmetry is intimately tied with Lorentz invariance. A massive particle of spin $s$ has $2s+1$ distinct spin states which mix under a general Lorentz transformation. The helicity of a massless particle, on the other hand, is frame invariant. Indeed, for free particles one can write down local fields which create or destroy just a single helicity state. For spin 1/2 Fermions coupled minimally to gauge fields, their helicity remains naively conserved.

In one space dimension the roles of left and right handed helicities are replaced by left and right moving particles. Since an observer cannot go faster than light, he can never overtake a massless particle and a right mover will be so in all Lorentz frames.

The fact that Lorentz invariance is crucial here provides another warning that chiral symmetry on the lattice will be difficult. Indeed, lattice formulations inherently violate the usual space time symmetries. Chiral issues should only be expected to be useful for states of low energy which do not see the underlying lattice structure.

While separate phase rotations of left and right handed massless Fermions give a formal symmetry of a continuum gauge theory, this is broken by the anomalies mentioned above. In particular, there is the famous triangle diagram where a virtual Fermion loop couples an axial vector current to two vector currents. In two space time dimensions the analogous problem arises with a simple bubble diagram connecting a vector and an axial vector current.

The fact that the anomaly must exist can be intuitively argued in analogy with band theory in solid state physics. With massive Fermions the vacuum has a Fermi level midway in a gap between the filled Dirac sea and a continuum of positive energy particle states. This represents an insulator. As the mass is taken to zero, the gap closes and the vacuum becomes a conductor. External gauge fields applied to this conductor can induce currents. For a specific example, consider a one space dimensional world compactified into a ring. A changing magnetic field through this ring will induce currents, changing the relative number of left and right moving particles. Without the anomaly, transformers would not work.

In this problem, physics should be periodic in the amount of flux through the ring. This is a two dimensional analog of the periodicity of four dimensional non-abelian gauge theories as one passes through topologically non-trivial configurations [8]. The latter case with the standard model gives rise to a non-conservation of the baryon current [4].

With the one dimensional ring, the strength of the flux characterizes the phase that a charged particle acquires in running around the ring. As this net phase adiabatically increases, the individual Fermionic energy levels shift monotonically. As one adds another unit of flux through the ring, one filled right moving level from the Dirac sea moves, say, up to positive energy, while one empty left moving level drops into the sea, leaving a hole. This induces a net current carried by a right moving particle and a left moving antiparticle. This way of visualizing how the anomaly works was nicely discussed some time ago [9].

### 3. The doubling problem

The essence of the lattice doubling problem already appears with the simplest Fermion Hamiltonian in one space dimension

$$H = iK \sum_j a^\dagger_{j+1} a_j - a^\dagger_j a_{j+1}. \tag{1}$$

Here $j$ is an integer labeling the sites of an infinite chain and the $a_j$ are Fermion annihilation operators satisfying standard anticommutation relations

$$\left[a_j, a^\dagger_k\right]_+ \equiv a_j a^\dagger_k + a^\dagger_k a_j = \delta_{j,k}. \tag{2}$$

The bare vacuum $|0\rangle$ satisfies $a_j|0\rangle = 0$. This vacuum is not the physical one, which contains a filled Dirac sea. I refer to $K$ as the hopping parameter. Energy eigenstates in the single Fermion sector

$$|\chi\rangle = \sum_j \chi_j a^\dagger_j |0\rangle \tag{3}$$

can be easily found in momentum space

$$\chi_j = e^{iqj} \chi_0. \tag{4}$$

where $0 \leq q < 2\pi$. The result is

$$E(q) = 2K \sin(q). \tag{5}$$

The physical vacuum has all negative energy states filled to form a Dirac sea. Particles are represented by excitations on this vacuum.

If I consider a Fermionic wave packet produced from a superposition of states carrying small momentum $q$, then, since the group velocity $dE/dq$ is positive in this region, the packet will move to the right. On the other hand, a wave packet produced from momenta in the vicinity of $q \sim \pi$ will be left moving. The essence of the Nielsen Ninomiya theorem [10] is that we must have both types of excitation. The periodicity in $q$ requires the dispersion relation to have an equal number of zeros with positive and negative slopes.

The recent attempts to circumvent this result add to the spectrum an infinite number of additional states at high energy[11]. The idea is to have a mode with $E = 2K \sin(q)$ still exist at small $q$, but then become absorbed in an infinite band of states before $q$ reaches $\pi$. If the band is truly infinite, then the extra state does not have to reappear as the momentum increases to $2\pi$. In the domain wall picture, this infinite tower of states is represented by a flow into an extra dimension [5,6].

## 4. The Wilson approach

In this section I review Wilson's scheme for adding a non-chirally symmetric term to remove the doublers appearing in a naive lattice transcription of the Dirac equation. I do this in some detail because the general behavior of the Wilson-Fermion Hamiltonian will be central to the later construction of surface modes. To keep the discussion simple, I work in one dimension with a two component spinor

$$\psi = \begin{pmatrix} a \\ b \end{pmatrix}. \tag{6}$$

The most naive lattice Hamiltonian begins with the simple hopping case of Eq. (1) and adds in the lower components and a mass term to mix the upper and lower components

$$\begin{aligned} H = iK \sum_j a^\dagger_{j+1} a_j - a^\dagger_j a_{j+1} - b^\dagger_{j+1} b_j + b^\dagger_j b_{j+1} \\ + M \sum_j a^\dagger_j b_j + b^\dagger_j a_j. \end{aligned} \tag{7}$$

Introducing Dirac matrices

$$\gamma_0 = \begin{pmatrix} 0 & 1 \\ 1 & 0 \end{pmatrix}, \quad \gamma_1 = \begin{pmatrix} 0 & -1 \\ 1 & 0 \end{pmatrix} \tag{8}$$

and defining $\overline{\psi} = \psi^\dagger \gamma_0$, I can write the Hamiltonian more conventionally as

$$H = \sum_j iK(\overline{\psi}_{j+1} \gamma_1 \psi_j - \overline{\psi}_j \gamma_1 \psi_{j+1}) + M \sum_j \overline{\psi}_j \psi_j. \tag{9}$$

As before, the single particle states are easily found by Fourier transformation and satisfy

$$E^2 = 4K^2 \sin^2(q) + M^2 \tag{10}$$

Again, the negative energy sea is to be filled.

Naive chiral symmetry is implemented through distinct phase rotations for the upper and lower components of $\psi$. The mass term mixes these components and opens up a gap in the spectrum. The doublers at $q \sim \pi$, however, are still with us.

To remove the degenerate doublers, I make the mixing of the upper and lower components momentum dependent. A simple way of doing this was proposed by Wilson [12]. For this I add one more term to the Hamiltonian

$$\begin{aligned} H = & iK \sum_j a^\dagger_{j+1} a_j - a^\dagger_j a_{j+1} - b^\dagger_{j+1} b_j + b^\dagger_j b_{j+1} \\ & + M \sum_j a^\dagger_j b_j + b^\dagger_j a_j \\ & -rK \sum_j a^\dagger_j b_{j+1} + b^\dagger_j a_{j+1} + b^\dagger_{j+1} a_j + a^\dagger_{j+1} b_j \\ = & \sum_j K(\overline{\psi}_{j+1}(i\gamma_1 - r)\psi_j - \overline{\psi}_j(i\gamma_1 + r)\psi_{j+1}) + \sum_j M\overline{\psi}_j \psi_j. \end{aligned} \quad (11)$$

Now the spectrum satisfies

$$E^2 = 4K^2 \sin^2(q) + (M - 2rK \cos(q))^2. \quad (12)$$

Note how the doublers at $q \sim \pi$ are increased in energy relative to the states at $q \sim 0$. The physical particle mass is now $m = M - 2rK$ while the doubler is at $M + 2rK$.

The hopping parameter has a critical value at

$$K_{crit} = \frac{M}{2r} \quad (13)$$

At this point the gap in the spectrum closes and one species of Fermion becomes massless. The Wilson term, proportional to $r$, still mixes the $a$ and $b$ type particles; so, there is no exact chiral symmetry. Nevertheless, in the continuum limit this represents a candidate for a chirally symmetric theory. Beforehand, as discussed in Ref. [13], chiral symmetry does not provide a good order parameter.

A difficulty with this approach is that gauge interactions will renormalize the parameters. To obtain massless pions one must finely tune $K$ to $K_{crit}$, an *a priori* unknown function of the gauge coupling. Despite the awkwardness of such tuning, this is how numerical simulations with Wilson quarks generally proceed. The hopping parameter is adjusted to get the pion mass right, and one hopes for the remaining predictions of current algebra to reappear in the continuum limit.

## 5. Supercritical $K$ and surface modes

The case of $K$ exceeding the critical value $M/2r$ is rarely discussed but quite interesting nevertheless. Aoki and Gocksch [13] have argued that as one passes through this point with gauge fields present, there occurs a spontaneous breaking of parity.

Restricting ourselves to the free Fermion case for the time being, interesting things happen here for supercritical $K$ as well. As the band closes and reopens with increasing $K$, the positive energy particles and the negative energy Dirac sea couple strongly. A similar situation was studied some time ago by Shockley [14], who observed that if the system is finite with open walls, then two discrete levels leave the bands and emerge bound to the ends of the system.

As the volume of the system goes to infinity, particle-hole symmetry forces these surface levels to go to exactly zero energy. In a finite box, the wave functions have exponential tails away from the walls, mixing the states and in general giving them a small residual energy.

A general result [1] is that there exists such a state bound to any interface separating a region with $K > K_{crit}$ from a region with $K < K_{crit}$. In Ref. [5], Kaplan uses $M = 2Kr + m\epsilon(x)$. I prefer to consider here the simpler approach of Shamir [15] and take $K = 0$ on one side, giving modes on an open surface.

In the later discussion of the anomaly in terms of currents into an extra dimension, it will always be a flow into a region of supercritical hopping. This should be contrasted with the continuum discussion of Ref. [16], where the flow is symmetric about the defect. This symmetry appears, however, to be regulator dependent [17]. For example, with a Pauli-Villars regulator, the relative sign of the Fermion to the regulator masses controls the direction of flow.

Following the usual procedure of filling half the states for the Dirac sea, we see that there is an ambiguity with the last Fermion, which could go into either of the degenerate surface modes. If I imagine coupling the Fermions to, say, a $U(1)$ gauge field, then this last Fermion will be a source of a background electric field which will run to the hole state on the opposite wall. This is the physical origin of the parity breaking proposed in Ref. [13]. In the continuum limit the vacuum should be equivalent to that of the massive Schwinger model with a half unit of background electric flux. The physics of this model in the continuum was extensively discussed in Ref. [18].

## 6. Extra dimensions

As the system size goes to infinity, particle-hole symmetry naturally forces the the surface modes to zero energy. This behavior forms the basis for a lattice approach to chiral Fermions. The picture of Kaplan [5] is to reinterpret the coordinate labeled by $j$ in the above discussion as an extra dimension beyond the usual ones of space and time. Our physical world then exists on a four dimensional interface, with the light quarks and leptons being the above surface modes.

To be concrete, consider adding $D$ space dimensions to the above Hamiltonian, where for the following $D$ will either be 1 or 3. For simplicity I will take $L^D$ space sites and use antiperiodic boundary conditions for each of these dimensions. The extra dimension, which I refer to as the fifth, has $L_5$ sites and open boundaries. I take the same hopping and Wilson parameters in each of the dimensions, including the fifth, although this is not essential.

The Dirac matrices $\gamma_\mu$ satisfy the usual

$$[\gamma_\mu, \gamma_\nu]_+ = 2g_{\mu\nu}. \tag{14}$$

I define $\gamma_5 = i\gamma_0\gamma_1\gamma_2\gamma_3$ for $D = 3$ and $\gamma_5 = \gamma_0\gamma_1$ for $D = 1$. I take $\gamma_0$ and $\gamma_5$ to be Hermitian, while the spatial $\gamma$ matrices are anti-Hermitian. The Hamiltonian I am led to is then

$$\begin{aligned} H = \sum_{\mathbf{n},j} \Big( & K\overline{\psi}_{\mathbf{n},j+1}(\gamma_5 - r)\psi_{\mathbf{n},j} - K\overline{\psi}_{\mathbf{n},j}(\gamma_5 + r)\psi_{\mathbf{n},j+1} \\ & + \sum_{a=1}^{D} (K\overline{\psi}_{\mathbf{n}+\mathbf{e}_a,j}(i\gamma_a - r)\psi_{\mathbf{n},j} - K\overline{\psi}_{\mathbf{n},j}(i\gamma_a + r)\psi_{\mathbf{n}+\mathbf{e}_a,j}) \\ & + M\overline{\psi}_{\mathbf{n},j}\psi_{\mathbf{n},j} \Big). \end{aligned} \tag{15}$$

Here $\mathbf{n}$ denotes the spatial sites, $j$ the extra coordinate, and $\mathbf{e}_a$ is the unit vector in the positive $a$'th direction.

The use of antiperiodic boundary conditions makes it simple to go to momentum space for the spatial coordinates. Denoting the components of the momentum by $q_a$, I write

$$\psi_{\mathbf{q},j} = \frac{1}{L^{D/2}} \sum_{\mathbf{n}} e^{-i\mathbf{q}\cdot\mathbf{n}} \psi_{\mathbf{n},j}. \tag{16}$$

Each component of the momentum takes discrete values from the set $(2k + 1)\pi/L$ where $k$ runs from, say, $0$ to $L - 1$. This makes the Hamiltonian block diagonal, with each value for $\mathbf{q}$ representing a separate block. In this way the Hamiltonian reduces to

$$\begin{aligned} H = \sum_{\mathbf{q},j} \Big( & K\overline{\psi}_{\mathbf{q},j+1}(\gamma_5 - r)\psi_{\mathbf{q},j} - K\overline{\psi}_{\mathbf{q},j}(\gamma_5 + r)\psi_{\mathbf{q},j+1} \\ & + \sum_a 2K\sin(q_a)\overline{\psi}_{\mathbf{q},j}\gamma_a\psi_{\mathbf{q},j} \\ & + (M - 2Kr\sum_a \cos(q_a))\overline{\psi}_{\mathbf{q},j}\psi_{\mathbf{q},j} \Big). \end{aligned} \tag{17}$$

Modes bound to the surface in the fifth direction exist whenever $K$ exceeds the critical value

$$K_{crit} = M/2r - K\sum_a \cos(q_a). \tag{18}$$

Note how this critical value now depends on the spatial momentum. Appropriately choosing $M$, I can have the surface states exist for small $q$, but have them disappear when any component of $q_a \sim \pi$. This avoids the doublers [19]. Specifically, when the hopping is direction independent, I want (assuming $K$, $r$, and $M$ are all positive)

$$(D - 1)K < M/2r < (D + 1)K. \tag{19}$$

The above discussion shows that on a single surface I have an elegant lattice theory for a low energy chiral Fermion. I would now like to add gauge fields. Here I adopt the attitude that I do not want a lot of new degrees of freedom, and follow Ref. [11] in regarding the extra dimension as a flavor space. In particular, I do not put gauge fields in the fifth dimension, and the physical gauge fields are independent of this dimension.

While this approach has the advantage of preserving an exact gauge invariance and not introducing lots of unwanted fields, it has the disadvantage that both walls are coupled equally to the gauge field. Thus, even when the size of the fifth dimension approaches infinity, the opposite chirality Fermions do not decouple. The main thing that has been accomplished so far is to find a theory of Fermions coupled in a vectorlike manner, without any doublers, and with a natural way to take the Fermion masses to zero.

## 7. The anomaly and rotating eigenvalues

One of the nice features of this formulation is how the chiral anomaly appears as a flow of Fermionic states in the extra dimension. The basic scenario was discussed in a somewhat different context by Callan and Harvey [16]. They consider a vector theory, whose mass term has a domain wall shape in an extra dimension, and show that it has a chiral zeromode living on the wall. The anomalous gauge current generated by this state has to be cancelled in the underlying $2n+1$ dimensional theory since that world is anomaly free. Indeed, the massive modes contribute to the low energy effective action a piece representing the flow of charge into (or out of) the wall from the extra dimension. When calculated far from the wall, it cancels the anomalous contribution. In the $U(1)$ case in 2+1 dimensions this was recently explicitly checked on the lattice with both Kaplan's and Shamir's formulations [20]. Indeed the cancellation is valid even close to the wall [21]. Therefore, what on the interface looks like an anomaly is the flow of charge into the extra dimension and the role of the heavy modes is to carry that charge.

The above picture was studied in some detail in Ref. [22]. Since opposite chirality partners live on opposite walls, the charge has to be transported through the extra dimension. In the adiabatic limit of slowly varying gauge fields, the time evolution is a continuous change of one particle states. As one passes through an "instanton" configuration the low energy states at the lattice ends change energy without substantially changing their position in the extra dimension. The same is true for the very high energy states, residing deep in the lattice interior.

However, the surface states with energies close to the cutoff are very sensitive to the applied field. When the energy of such a level rises towards the bottom of the band of plane waves flowing in the extra coordinate, the wave function penetrates increasingly deeply into this dimension. At the same time, another level from the interior lowers its energy and flows towards the opposite wall. This is also true for levels with corresponding negative energies; they just move in the opposite direction.

In this way we see how the heavy modes right at the cutoff carry the charge on and off the surfaces. With a gauge field applied to the physical vacuum with all negative energy states filled, these "flying states" are responsible for what appears to be the gauge anomaly on the surfaces.

## 8. Weak interactions, mirror Fermion model

With an exact gauge invariance and a finite size for the extra dimension, the surface models are inherently vectorlike. The Fermions always appear with both chiralities, albeit separated in the extra dimension. However, experimentally we know that only left handed neutrinos couple to the weak bosons. In this section I discuss one way to break the symmetries between these states, resulting in a theory with only one light gauged chiral state. Here I keep the underlying gauge symmetry exact, but do require that the chiral gauge symmetry be spontaneously broken, just as observed in the standard model. The picture also contains heavy mirror Fermions. If anomalies are not cancelled amongst the light species, these heavy states must survive in the continuum limit. It remains an open question when anomalies are properly cancelled whether it might be possible to drive the heavy mirror states to arbitrarily large mass.

I start by considering two separate species $\psi_1$ and $\psi_2$ in the surface mode picture. However, I treat these in an unsymmetric way. For $\psi_1$ I use the previous Hamiltonian. For $\psi_2$ I change the sign of all terms proportional to $\gamma_5$. On a given wall, the surface modes associated with $\psi_1$ and $\psi_2$ will then have opposite chirality.

Now I introduce the gauge fields. Since I want to eventually couple only one-handed neutrinos to the vector bosons, consider gauging $\psi_1$ but not $\psi_2$. Indeed, at this stage $\psi_2$ represents a totally decoupled right handed Fermion on one wall. I still have a mirror situation on the opposite wall, consisting of a right handed gauged state and a left handed decoupled Fermion.

The next ingredient is to spontaneously break the gauge symmetry, as in the standard model, by introducing a Higgs field $\phi$ with a non-vanishing expectation value. I can use this field to generate masses as in the standard model by coupling $\psi_1$ and $\psi_2$ with a term of the form $\bar\psi_1 \psi_2 \phi$.

The new feature is to allow the coupling of the Higgs field to depend on the extra coordinate. In particular, let it be small or vanishing on one wall and large on the other. The surface modes are then light on one wall and heavy on the other.

This model is closely related to the proposal in [23], where the gauge field is suddenly shut off in the interior of the fifth dimension, and gauge invariance is restored via a Higgs field. Folding that lattice in half around this shut off point reduces it to the picture presented here.

As in other mirror Fermion models [24], triviality arguments suggest that there might exist bounds on the mass of the heavy particles. This is certainly expected to be the case where the light Fermions alone give an anomalous gauge theory, in which case I expect the mirror particles cannot become much heavier than the vector mesons, i.e. the $W$.

It is conceivable that the restrictions on the mirror Fermion masses are weaker when anomalies cancel amongst the light states. In this case there is no perturbative need for the heavy states, and perhaps they can be driven to infinite mass in the continuum limit. This is a rather speculative desire, but if possible would give a candidate for a lattice discretization of the standard model.

Unfortunately, the model as it stands does not lead to baryon number violation [25]. The anomaly will involve a tunnelling of baryons from one wall to the opposite, where they become mirror baryons. Even if these extra particles are heavy, the decay can only occur through mixing with the ordinary particle states. In this sense, the mirror particles still show their presence in low energy physics. This further hints that the mirror Fermions might not be removable in the continuum limit.

Another speculative proposal is to use the right handed mirror states in some way as observed particles. Indeed, the world has left handed leptons and right handed antibaryons. Any simple extension of this idea to a realistic model must unify these particles [26]. On the other hand, the fact that the anomalies are canceled between different representations of the $SU(3)$ of strong interactions may preclude such options.

## 9. Conclusions

The use of Shockley surface states may provide the basis for a theory of chiral Fermions. For strong interaction physics this yields an elegant formulation where the massless limit for the quarks is quite natural.

In this picture the anomaly appears as a flow into the extra dimension. For anomaly free currents, the net flow in this dimension cancels, and I expect the predictions of current algebra to arise naturally. On the other hand, the symmetries for singlet axial currents are strongly broken by this flow. This presumably precludes the need for a corresponding Goldstone boson and solves the $U(1)$ problem.

Several questions remain before we have a theory of the weak interactions on the lattice, where the gauge fields are to be coupled to chiral currents. One approach leads to a theory with mirror Fermions on the opposing walls of the system. In a spontaneously broken theory these extra states can be given different masses. Whether they can be driven to infinite mass in the continuum limit presumably depends on whether all necessary chiral anomalies have been cancelled.

The difficulties in formulating chiral theories with a fundamental non-perturbative cutoff, such as the lattice, hints that there may be a deeper hidden message. Perhaps mirror fermions must exist at a few times the $W$ mass and we should be looking for them. Note also that spontaneous breaking of the gauge theory is central to that approach, hinting that perhaps the only consistent chiral theories are spontaneously broken.